\documentclass[reprint, pdftex, amsmath,amssymb, aps, prl, showpacs]{revtex4-1}

\usepackage{graphicx}
\usepackage{dcolumn}
\usepackage{bm}
\usepackage{hyperref}
\usepackage{color}

\DeclareGraphicsExtensions{.pdf, .png, .jpg}

\newcommand{\drain}{\ensuremath{|^1 \textrm{S}_0\rangle \,-\,|^3 \textrm{P}_1 \rangle \:}}

\begin{document}

\preprint{APS/123-QED}

\title{Observation of Motion Dependent Nonlinear Dispersion with Narrow Linewidth Atoms in an Optical Cavity}

\author{Philip G. Westergaard$^{1,2}$} 
 \email{pgw@dfm.dk}
\author{Bjarke T. R. Christensen$^1$}%
\author{David Tieri$^3$}%
\author{Rastin Matin$^1$}%
\author{John Cooper$^3$}%
\author{Murray Holland$^3$}%
\author{Jun Ye$^3$}%
\author{Jan W. Thomsen$^1$}%

\affiliation{%
$^1$Niels Bohr Institute, University of Copenhagen; Blegdamsvej 17, 2100 Copenhagen, Denmark\\
$^2$Danish Fundamental Metrology; Matematiktorvet 307, 1. sal, 2800 Kgs. Lyngby, Denmark\\
$^3$JILA, National Institute of Standards and Technology and University of Colorado, Boulder, CO 80309-0440, USA
}%

\begin{abstract}
As an alternative to state-of-the-art laser frequency stabilisation using ultra-stable cavities, it has been proposed to exploit the non-linear effects from coupling of atoms with a narrow transition to an optical cavity. Here we have constructed such a system and observed non-linear phase shifts of a narrow optical line by strong coupling of a sample of strontium-88 atoms to an optical cavity. The sample temperature of a few mK provides a domain where the Doppler energy scale is several orders of magnitude larger than the narrow linewidth of the optical transition. This makes the system sensitive to velocity dependent multi-photon scattering events (Dopplerons) that affect the cavity field transmission and phase. 
By varying the number of atoms and the intra-cavity power we systematically study this non-linear phase signature which displays roughly the same features as for much lower temperature samples. This demonstration in a relatively simple system opens new possibilities for alternative routes to laser stabilization at the sub 100 mHz level and superradiant laser sources involving narrow line atoms. The understanding of relevant motional effects obtained here has direct implications for other atomic clocks when used in relation with ultranarrow clock transitions.

\end{abstract}
\maketitle


State-of-the-art atomic clocks rely on highly coherent light sources to probe narrow optical transitions \cite{BenBloom,Targat_NatCom,Hinkley,Katori2005,Chou}. However, these clocks are limited by the frequency noise of the interrogation oscillator through the Dick effect \cite{Santarelli}. Only recently multi-atom optical clocks have surpassed single ion clocks in stability owing to the enhanced laser stability \cite{Nicholson_PRL,Hinkley,BenBloom}. Achieving a better stability has so far been hampered by thermal noise in the reference cavity used for laser stabilisation \cite{Kessler_2,Kessler,MartinScience}.
Recent proposals suggest an alternative approach to laser stabilisation \cite{Meiser2, MikeMartin, Meiser1} where atoms in an optical lattice are probed on the narrow clock transition inside an optical cavity. This brings non-linear effects into the system dynamics that could considerably enhance the spectral sensitivity and potentially lead to laser stability comparable to or better than the current state-of-the-art. However, for finite temperature samples of atoms the principal mechanisms that are relevant to this physical domain have not been investigated in detail.

\begin{figure*}[ht]
\includegraphics[width=0.70\textwidth]{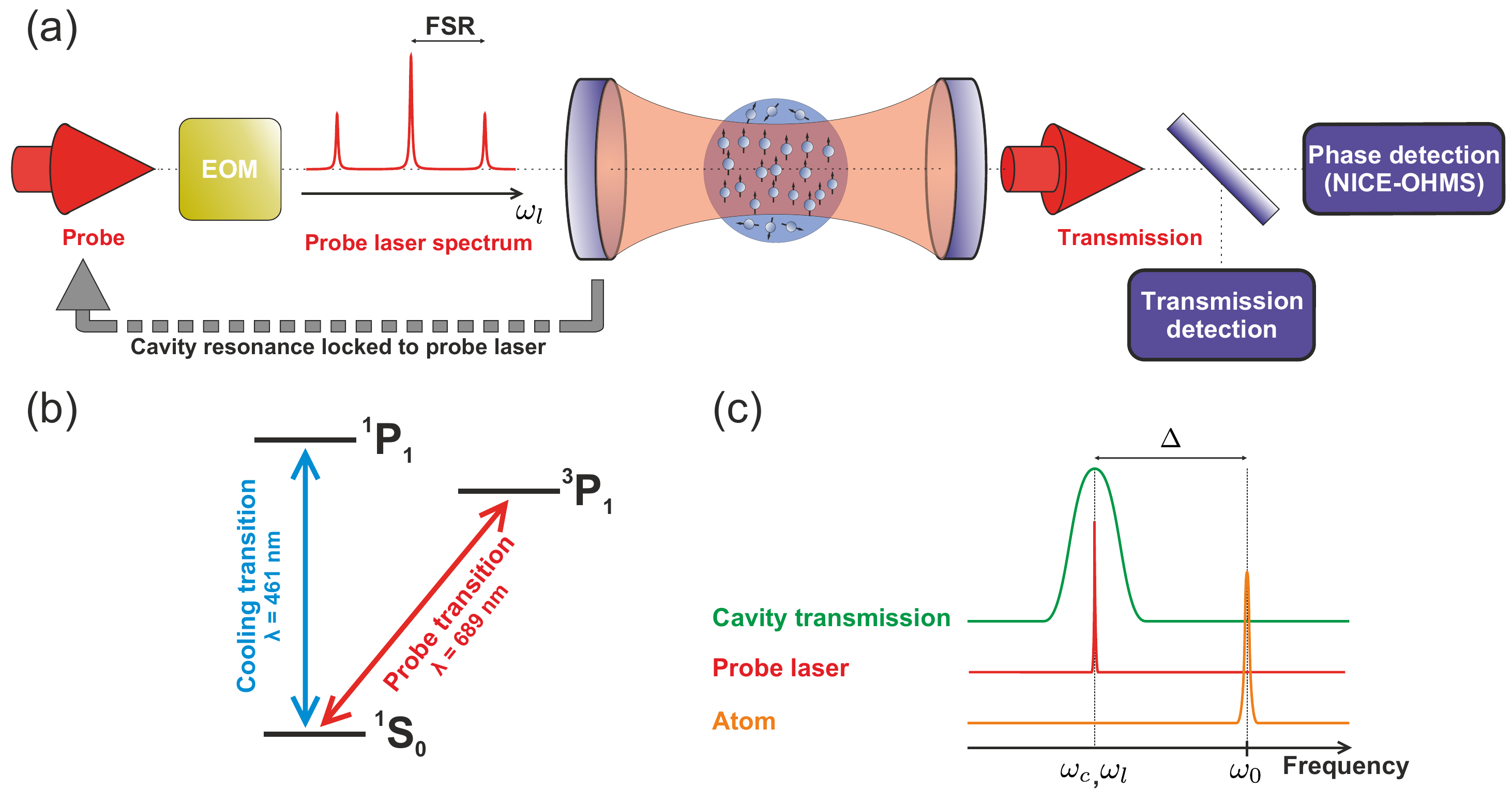}
\caption{\small (color online). (a) Experimental setup. A sample of cold atoms (MOT) is prepared inside a low finesse cavity ($F = 85$) which is held at resonance with the probe laser. We probe the atoms on the inter-combination line $|5s^{2} \, ^1 \textrm{S}_0 \rangle \,-\, | 5s5p \, ^3 \textrm{P}_1 \rangle$ at $689$ nm ($\Gamma/2\pi  = 7.6$ kHz). Both intensity and phase shift of the transmitted probe light are recorded. The phase is measured relative to the input field by employing cavity-enhanced heterodyne spectroscopy (NICE-OHMS). (b) Energy levels of the $^{88}$Sr atom and transitions relevant for this work.
(c) Relation between the spectral components in the experiment. The probe laser frequency $\omega_l$ (and consequently the cavity resonance $\omega_c$) is detuned a variable amount $\Delta$ with respect to the atomic resonance $\omega_0$.
\label{fig:Cavity}}
\end{figure*}

In such systems with highly non-linear phase response, \emph{a priori} unpredictable effects such as bi-stability \cite{MikeMartin} and the finite temperature of the atomic ensemble can change the phase response in an undesirable way, which could reduce the performance of the stabilisation scheme for all practical implementations. To achieve a better understanding of cavity-mediated effects with a narrow optical transition we have constructed a system with $^{88}$Sr atoms probed on the \drain transition at 689 nm inside an optical cavity (see Fig. \ref{fig:Cavity}). To capture the basic physics of the strong non-linear phenomena one can consider $N$ atomic dipoles strongly coupled to a single mode of the cavity field. The dipole moment associated with this narrow transition is around five orders of magnitude smaller than that for a typical dipole-allowed transition in an alkaline element. Also, at finite temperature only a small fraction of the atomic sample is probed due to Doppler broadening. Here, the role of the cavity is to enhance the weak interaction by order of the finesse of the cavity.

Experimentally we operate in the so-called "bad cavity" regime, where the atomic dipole decay rate is a factor of 1000 smaller than the cavity decay rate $\kappa$. In our experiment we use the $^{88}$Sr $|5s^{2} \, ^1 \textrm{S}_0 \rangle \,-\, | 5s5p \, ^1 \textrm{P}_1 \rangle$ transition at $461$ nm to cool and trap atoms in a magneto-optical trap (MOT). We load about $5\times 10^8$ atoms in the MOT at a temperature of $2-4$ mK inside an optical cavity prepared for light at 689 nm. The cavity waist of $w_0 = 500\; \mu$m ensures a good overlap with the MOT and negligible transit time broadening ($\sim 2$ kHz) compared to the natural line width ($\Gamma/2\pi = 7.6$ kHz) of the probe transition. The dimensionless number $C=C_0 N$, where $C_0 = 4g^2/\Gamma\kappa$ depends on the single-atom/cavity coupling constant $g$, is known as
the collective cooperativity and is a measure of how strong the coherent atom-cavity coupling is with respect to the dissipation channels.
In our configuration ($g/2\pi =590$ Hz, $\kappa/2\pi = 5.8$ MHz)  we are able to generate a collective cooperativity of about $C = 630$, thus placing our system in the regime of high collective cooperativity in the bad cavity limit, but outside the more restrictive CQED strongly coupled regime \cite{Vuletic, SupplInfo}.

Our experiment is operated in a cyclic fashion. We start each cycle preparing the atomic sample by loading a MOT inside the optical cavity. After loading we shut off the MOT beams and probe the atoms at 689 nm while recording both the intensity and phase shift of the transmitted probe light via two detectors (see Fig. \ref{fig:Cavity}). The total cycle time is typically around $0.5-1$ s. For the phase measurement we employ cavity-enhanced FM spectroscopy by using the so-called noise-immune cavity-enhanced optical-heterodyne molecular spectroscopy (NICE-OHMS) technique \cite{NICE-OHMS,NICE-OHMS2} (see Supplemental Information). This technique has a clear advantage over heterodyne signals generated, for example, from interferometric methods in terms of superior noise reduction and simplicity. During experiments we lock the cavity resonance to the 689 nm laser frequency using a H\"{a}nsch-Couillaud scheme \cite{Hansch-Couillaud}. The standing wave generated in the cavity will thus be present at all times while the 689 nm laser frequency is scanned.

In the limit of $T=0$ and for very low cavity field intensities several solutions exist for the steady-state intra-cavity field \cite{MikeMartin}. This is known as optical bi-stability, which would render the system unsuited for frequency stabilization. However, at finite temperatures when motional effects are included this picture changes. In this case, there is a critical temperature $T_{\textrm{crit}}$ above which only one solution for the steady-state intra-cavity field exists. For our parameters $T_{\textrm{crit}}$ is of the order of a few hundred nK while experiments are typically performed at mK temperatures.

\begin{figure}
\includegraphics[width=0.85\columnwidth]{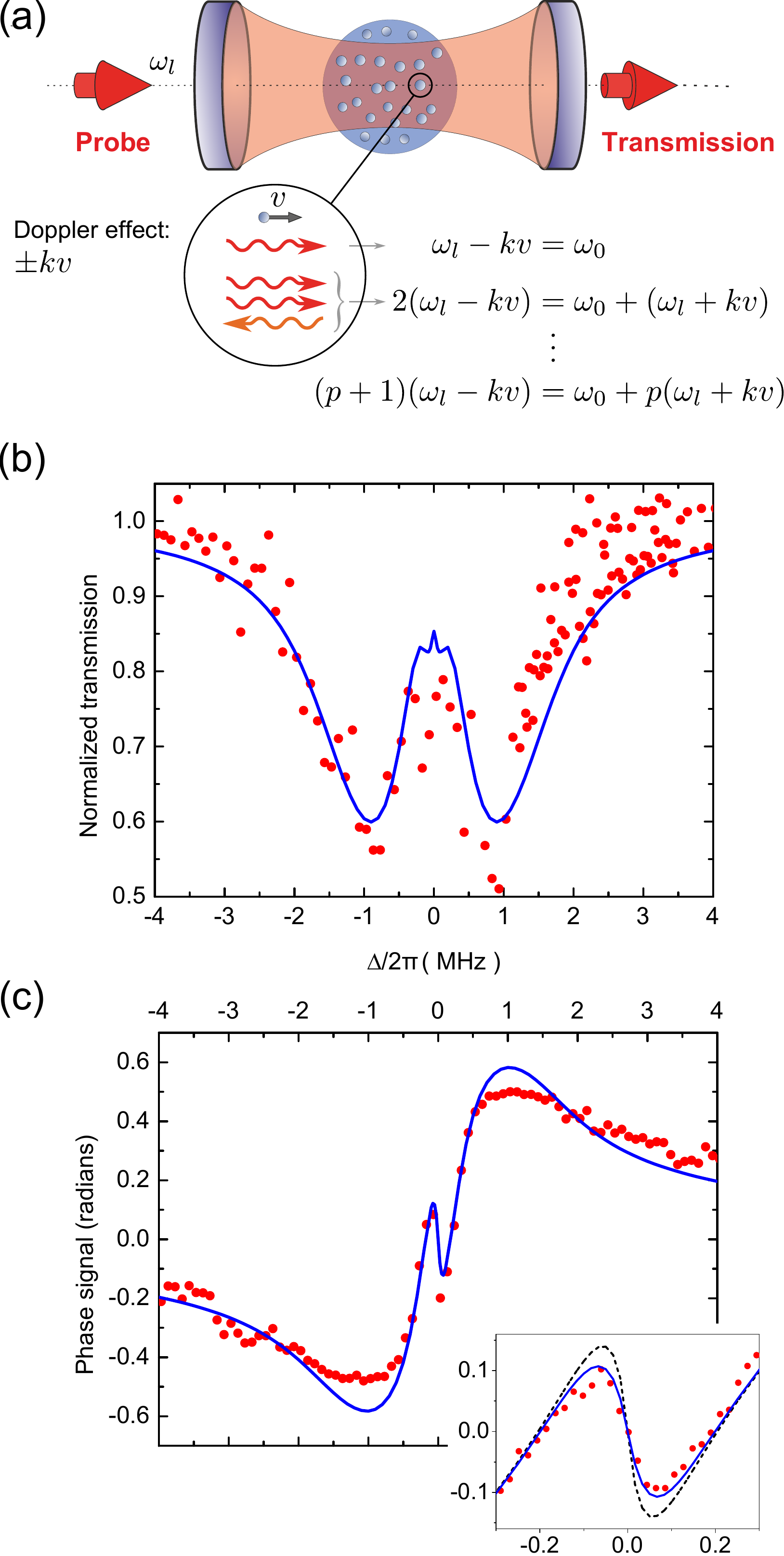}
\caption{\small (color online). (a) Illustration of the Doppleron multi photon processes that take place in our system. We consider a given atom with velocity component $v$ in the direction of the cavity axis. The first resonance condition (top equation) involves only one photon and corresponds to the usual Doppler effect. The next resonance involves two photons absorbed and one photon emitted, and so forth. (b, c) Typical frequency scan without any averaging across the atomic resonance for an input power of 975 nW and a total number of atoms in the MOT of $N= 4.4\cdot 10^8$. The data in (b) displays the transmission of the probe light through the cavity normalized to a signal with no atoms in the cavity. The data in (c) is the phase shift of the cavity-transmitted field obtained using the NICE-OHMS method. The solid lines are theoretical predictions based on our theoretical model which includes the Doppler effect and the spatial overlap of the thermal cloud (here with temperature $T= 2.3$ mK) with the cavity field. At maximum phase shift (around detunings of $\Delta \simeq \pm 1$ MHz) our detection system starts to saturate giving a slightly flatter appearance of the phase data. Inset: Zoom on central phase feature with similar experimental parameters (data identical to Fig. 5(a)). Here, we have included a theoretical plot that does not take the Dopplerons into account (black, dashed curve). The effect of the Dopplerons is readily apparent. Units on axes are the same as in (c).
\label{fig:Power transmitted}}
\end{figure}

The non-zero velocity of the atoms brings additional photon resonance phenomena into play, which changes the complex amplitude of the cavity field around the atomic resonance $\omega_0$. In the rest frame of an atom moving with velocity $v_j$ the atom experiences a bi-chromatic light field given by $\omega_+=\omega_l \left( 1+v_j/c \right)$ and $\omega_-=\omega_l \left( 1 - v_j/c \right)$, where $\omega_l$ is the laser frequency and $c$ is the speed of light. Resonant scattering events will take place if the atom is Doppler-tuned into resonance at $\omega_0$, e.g.,  $\omega_-=\omega_0$, such that the atom may absorb a photon from a given direction of the cavity field. Higher order resonances are also possible where, e.g., the atom absorbs two photons from one direction at $\omega_-$ and emits one photon in the other direction at $\omega_+$. Generally, the resonance condition for $p+1$ absorbed and $p$ emitted photons becomes $(p+1)\omega_- =\omega_0 + p\omega_+$ for $p = 0,1,2,...$ \cite{Tallet}. The process is illustrated in Fig. \ref{fig:Power transmitted}(a). These non-linear multi-photon scattering effects are known as Dopplerons and give rise to a series of velocity dependent resonances \cite{Stenholm}, which change the transmitted field amplitude around resonance.

In Fig. \ref{fig:Power transmitted}(b, c) we show typical results for a frequency scan across the \drain line resonance (red circles). The input power was $975$ nW corresponding to an average saturation parameter of $S_0= 618$. It is clear that the phase signal in Fig. \ref{fig:Power transmitted}(c) has a significantly higher signal-to-noise ratio (SNR = 70) than the transmitted power signal in Fig. \ref{fig:Power transmitted}(b) (SNR $\sim 4$), demonstrating the effectiveness of the NICE-OHMS technique. Currently, the factor limiting the signal-to-noise ratio of the phase signal is the shot-to-shot atom number fluctuations and residual amplitude modulation from the EOM. 

We model the dynamics of the system by a Hamiltonian describing the coherent time evolution of an ensemble of atoms, where each atom with a given velocity is coupled to a single mode of the optical cavity. Solving the corresponding optical Bloch equations yields the cavity-transmitted intensity and phase as a function of detuning, number of atoms, and temperature. Our model is also adapted to take into account the spatial extent of the cavity field and atomic density profile. The blue solid curves in Fig. \ref{fig:Power transmitted} (b, c) are the theoretical prediction based on the Hamiltonian presented in Supplemental Equation 2 \cite{SupplInfo}.
In our theoretical model we fix the number of atoms, laser input power, laser line width, cavity waist, and cavity finesse based on experimental values, but allow a scaling factor for the absolute phase. The temperature is allowed to vary in the range of 2-4 mK, in accordance with the experimental condition.

Considering the transmission in Fig. \ref{fig:Power transmitted}(b) we can identify three spectral features: (1) the broad ($\sim 3$ MHz wide) Doppler absorption feature consistent with the sample temperature of a few mK; (2) a central region ($\sim 1$ MHz wide) with enhanced transmission due to saturation, affected by the Doppleron resonances which lead to enhanced back-scattering (or reduced forward transmission), limiting the height of the saturated absorption peak; (3) finally, in the central region around zero velocity (i.e., on resonance), the Doppleron mechanism breaks down and the saturated absorption takes place again with increased transmission as a result.

The Dopplerons also have an effect in the phase signal (Fig. \ref{fig:Power transmitted}(c)), although the effect is negligible for large laser detunings corresponding to larger atom velocities. In the inset of Fig. \ref{fig:Power transmitted}(c) we zoom in on the phase of the central saturated absorption feature were we have plotted experimental data (with parameters corresponding to Fig. \ref{fig:number dependence}(a)) and theoretical curves without Dopplerons (black, dashed) and with Dopplerons (blue). Here, the effect of Dopplerons becomes clear and there is an observable effect on the phase signature which is a decrease in slope around resonance, showing consistency between our theoretical model and the experimental data. This decrease in slope is important in the determination of the frequency stability that is achievable using this system, since the stability depends inversely on square of the slope around resonance, and reducing the temperature further does not significantly improve this slope \cite{SupplInfo}.

\begin{figure}[t]
\includegraphics[width=0.5\textwidth]{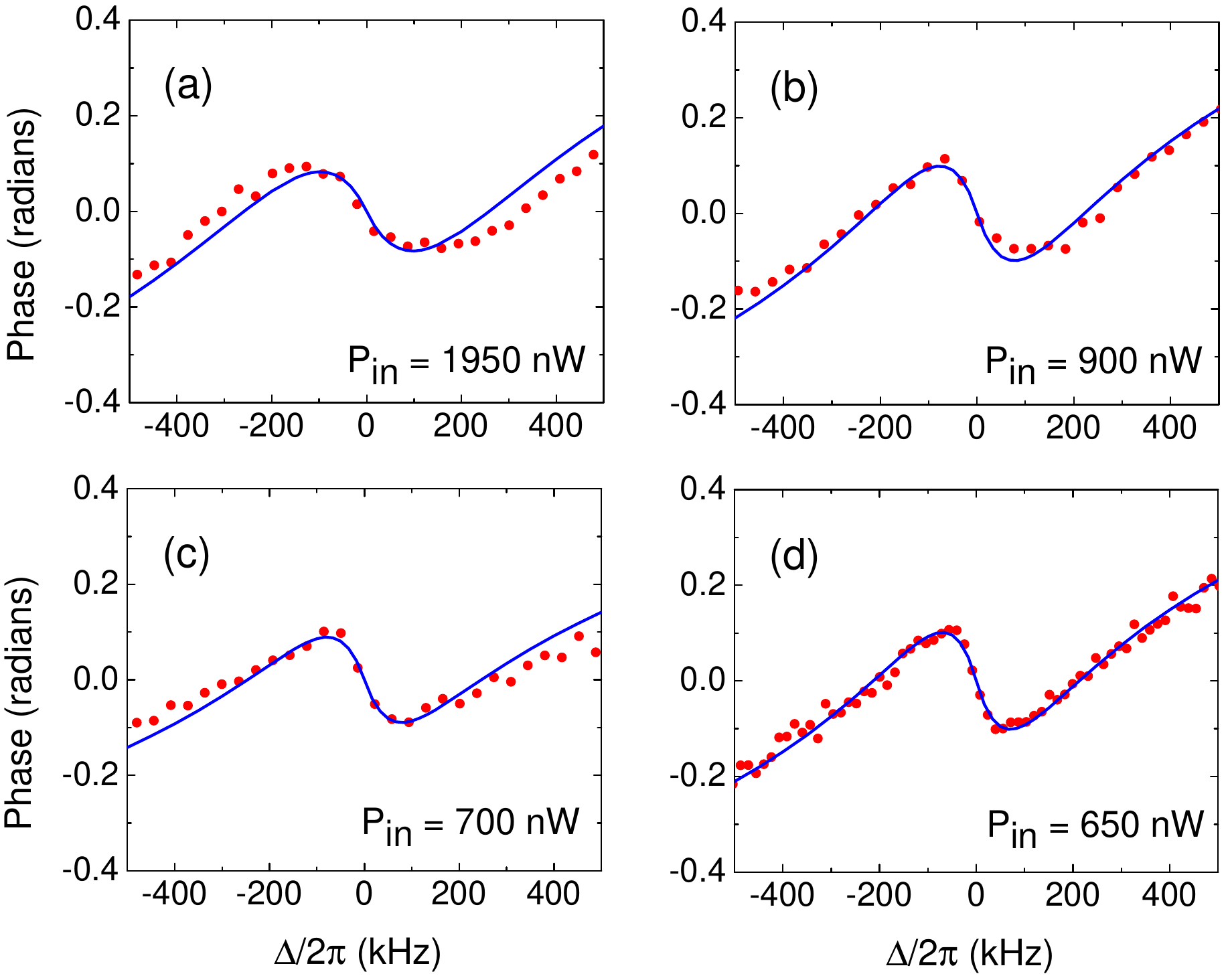}
\caption{\small (color online). Measured phase shift of the cavity-transmitted field when scanned across the atomic resonance.
The input probe laser power $P_{\textrm{in}}$ is progressively decreased from 1950 nW (a) to 650 nW (d). The number of atoms is about $N_{\textrm{cavity}} = 2.5\times 10^7$. Each point is an average of 3 data points. The solid lines are theoretical predictions based on our theoretical model.
\label{fig:Power dependence}}
\end{figure}

To evaluate and characterize our physical system experimentally and test it against the theoretical model we have mapped out the central phase feature as a function of probe input power with fixed atom number. In addition to a validation of the theoretical model this will provide an understanding of the behaviour and sensitivity of the phase signal to typical experimental variables relevant to, \emph{e.g.}, laser stabilisation. In Fig. \ref{fig:Power dependence} we show the phase signal for a fixed number of atoms as a function of laser detuning for different input powers in the range 650 - 1950 nW. For high input powers we strongly saturate the dipole and power broaden the central saturated absorption peak. As we gradually lower the input power, the power broadening is reduced leaving the central phase feature with a larger slope without reducing the signal-to-noise ratio.

\begin{figure}[h]
\includegraphics[width=0.5\textwidth]{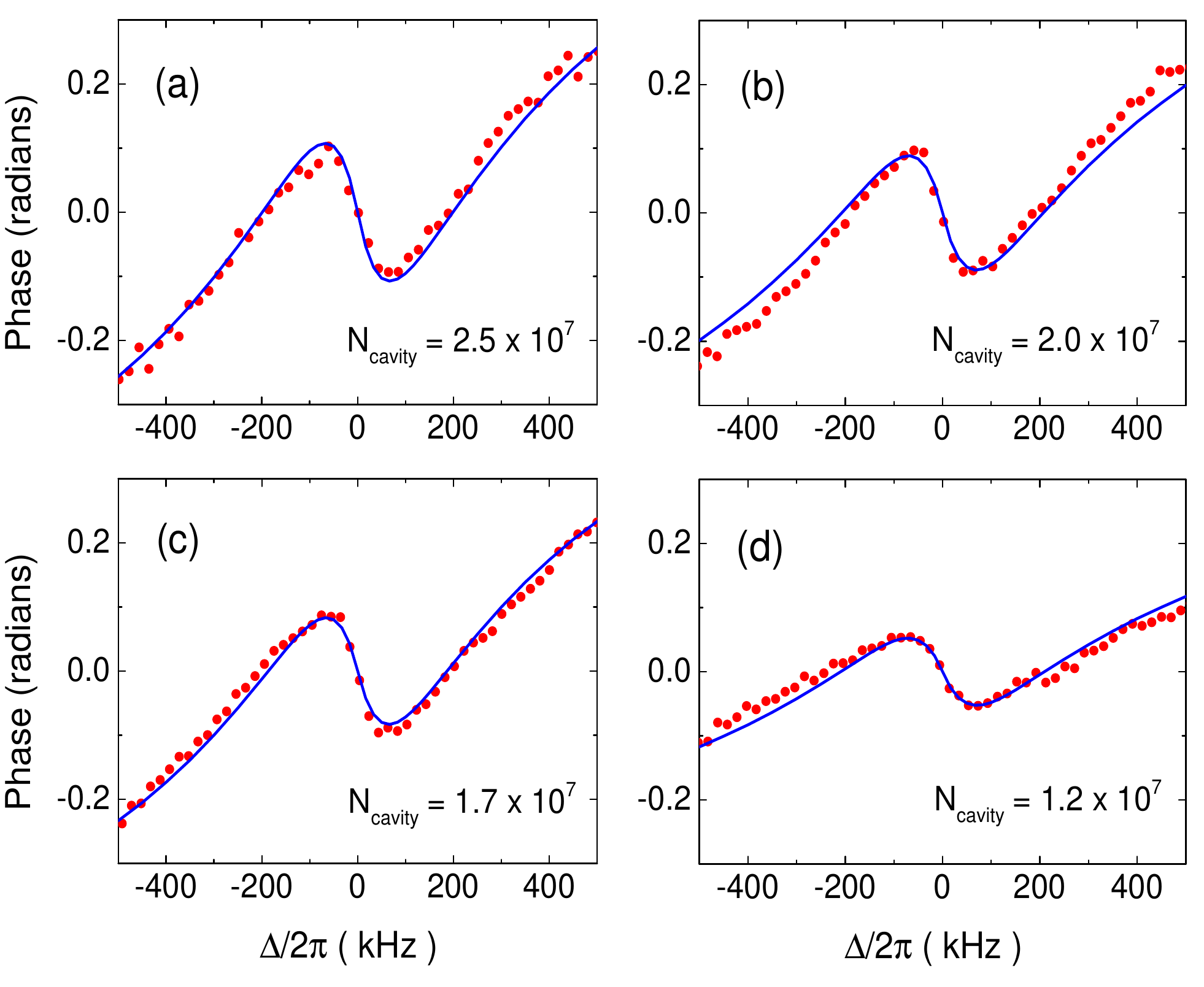}
\caption{\small (color online). Measured phase shift of the probe light when scanned across the atomic resonance. The number of atoms in the cavity is progressively decreased from from $2.5\times10^7$ (a) to $1.2\times 10^7$ (d). The input power used for all plots was 650 nW. Each point is an average of 3 data points. The solid lines are theoretical predictions based on our theoretical model. The central slope scales linearly with atom number.
\label{fig:number dependence}}
\end{figure}

Fig. \ref{fig:number dependence} (a) - (d) shows the evolution of the phase signal for fixed probe power as the number of atoms inside the cavity mode is changed from $ N_{\textrm{cavity}} = 2.5\times 10^7$ in (a) to $ N_{\textrm{cavity}}  =1.2\times 10^7$ in (d). 
We observe a strong dependence on atom number with increasing phase response and increasing slope on resonance for increasing atom numbers as expected, and the slope can straight forwardly be improved by increasing the number of atoms. However, our system is strongly non-linear and other optimal parameters, such as input power, for a given number of atoms, may not be trivially assigned to our experiment, but must be found numerically or experimentally.

Using the central phase slope for laser frequency locking, we estimate a shot noise limited line width of $1000\;$mHz based on our experimental parameters. This number can be improved by at least a factor of 20 with realistic improvements of the experimental parameters, \emph{e.g.} by optimizing the EOM modulation index (a factor 15), and increasing the atom number and the cavity finesse (both a factor 10), which would render the system comparable to state-of-the-art frequency stabilisation references \cite{Kessler, Bishof, Jiang, Millo} (see Supplemental Information for details).

In conclusion, we have constructed a system dominated by highly saturated multi-photon absorption with laser-cooled strontium atoms coupled to a low-finesse optical cavity. The transmission through the cavity is altered by thermal effects, but apart from a small decrease in slope, the central phase response of the atoms remains relatively immune to these effects while displaying a high signal-to-noise ratio (SNR) owing to the cavity and detection technique. The atomic phase signature was observed via cavity enhanced FM spectroscopy (NICE-OHMS) on the narrow optical \drain inter-combination line of $^{88}$Sr providing SNR of exceeding 7000 for one second of integration.
The understanding obtained here of the "bad cavity" physics lends promise to further development in this area, such as a new generation of frequency stabilisation \cite{Meiser2,MikeMartin} or superradiant laser sources \cite{Thompson, Maier}. Specifically, the physical understanding of a "warm" system (MOT temperature) obtained in his work will prove valuable when future atomic clocks and/or stable lasers will be operated under more noisy and compact environments, e.g. in vehicles and space crafts, where the size, ruggedness and convenience of the setup might dictate higher atomic temperatures than what is currently used for state-of-the-art systems. In this situation, this work will serve as an important piece of technical understanding for out-of-lab clocks employing warm atoms.

\begin{acknowledgements}
We would like to acknowledge support from the Danish research council and ESA contract No. 4000108303/13/NL/PA-NPI272-2012. DT, MH, and JY also wish to thank the DARPA QuASAR program, the NIST and the NSF for financial support.
\end{acknowledgements}

\clearpage 

\section*{Supplemental Information}\label{sec:Suppl}

\subsection*{Detection Scheme}
After loading the MOT for $0.5 - 1$ s we shut off the MOT beams and wait for $100\; \mu$s before probing the atoms for $100\; \mu$s. The probe laser beam is split up after the cavity and sent to two detectors which allow for a low bandwidth  detection ($50$ MHz) of the transmitted power and a high bandwidth detection ($2000$ MHz) for the phase measurement.
The NICE-OHMS phase signal is obtained by generating Fourier sidebands in the probe light at exactly the free spectral range (FSR) of our cavity at $500$ MHz using an EOM. The carrier is tuned close to the atomic resonance and will experience a phase shift due to the atoms. The sidebands, on the other hand, are far off resonance and will not be influenced by the atoms. They will, however, still be transmitted through the cavity since they are displaced by exactly one FSR, and experience similar technical noise and low-frequency amplitude noise as the carrier. Demodulating the heterodyne beat signal between the carrier and the sidebands after the high bandwidth detector will result in a signal proportional to the phase shift induced by the atoms at a significantly reduced noise level.

The transmission is measured by normalising the signal from the slow detector obtained just after the MOT is shut off with a signal obtained after waiting an additional 50 ms for all the atoms to have left the cavity. To minimize frequency fluctuations of the 689 nm laser we lock the laser
to a high finesse ULE cavity ($F\sim 8000$) using the Pound-Drever-Hall method \cite{PDH}. At 1 ms of integration time the laser has a linewidth of about $600$ Hz.

We monitor the number of atoms though the atomic absorption (low bandwidth detector) and indirectly through the MOT fluorescence measured by a photomultiplier. Typically, we trap about 5$\times$ $10^8$ atoms in the MOT while only $N_{\textrm{cavity} }=2\times 10^7$ atoms overlap with the cavity volume. The number of atoms is tuned by changing the loading time of the MOT.

\subsection*{Experimental Parameters}
By optical alignment of the MOT beams the trap center is off-set from the quadrupole magnetic field zero such that atoms experience a constant magnetic field of about 1 mT aligned parallel to the probe light polarization. This provides a local quantization axis and ensures that we only probe the $\pi$-transition; the $\sigma ^\pm $ transitions are shifted by tens of MHz. The mirrors for the 689 nm cavity are attached outside the view ports of the vacuum chamber, limiting the finesse of the cavity to $F = 85$ and the cavity length to $L = 30$ cm.

Our atom-cavity parameters are characterized by the single atom cooperativity $C_0 = \frac{6}{\pi^3}F\frac{\lambda^2}{w_0^2}= 3\cdot10^{-5}$, the atom-light coupling $g_0 =$ 590.6 Hz, the cavity decay time constant $\kappa$ and the atomic decay time constant $\Gamma$. Here $w_0$ is the cavity waist and $\lambda$ the transition wavelength. The collective coupling parameter is given by $g = g_0 \sqrt{N_{\textrm{cavity}}} $ where $N_{\textrm{cavity}}$ refers to the total number of atoms in the cavity ($N_{\textrm{cavity}}=2\times 10^7$). Our set of parameters is given by $(g,\kappa,\Gamma) = 2\pi\times (2.6\; \textrm{MHz},\; 5.8\; \textrm{MHz},\; 7.6 \;\textrm{kHz}$).
The collective cooperativity $C=C_0 N_{\textrm{cavity}}$ is for our system about $C=630$, thus placing our system in the the regime of high collective cooperativity (which requires $C\gg 1$) in the so-called bad cavity limit (the cavity linewidth $\kappa/2\pi = 5.8$ MHz is about a factor of 800 larger than the atomic linewidth $\Gamma/2\pi = 7.6$ kHz), but not in the CQED strong collective coupling regime which requires $g>>\Gamma$ and $g>>\kappa$ \cite{Vuletic}.  

\subsection*{Theoretical Model}
We model our system as a collection of 2-level atoms inside a single
mode optical cavity, using a Born-Markov master equation to describe
the open quantum system,
\begin{eqnarray}
  \frac{d}{dt} \hat{\rho} = \frac{1}{i \hbar} \left[ \hat{H}, \hat{\rho} \right] + \hat{\mathcal{L}}\left[ \hat{\rho} \right],
\label{ME1}
\end{eqnarray}
where,
\begin{equation}
\label{Hamiltonian}
\begin{split}
	\hat{H} \, = \, &\frac{\hbar \Delta}{2} \sum_{j=1}^{N} \hat{\sigma}^{z}_{j}
  + \hbar \eta \left( \hat{a}^{\dagger} + \hat{a}  \right) \\
  &+ \hbar  \sum_{j=1}^{N} g_j(t) \left( \hat{a}^{\dagger} \hat{\sigma}^{-}_{j}
    + \hat{\sigma}^{+}_{j} \hat{a} \right)\,.
	\end{split}
\end{equation}
The Hamiltonian $H$ describes the coherent evolution of the coupled
atom cavity system in an interaction picture which rotates at the
frequency of the cavity, and $\Delta = \omega_a - \omega_c$ is the atom-cavity detuning. The
Pauli spin matrices are $\hat{\sigma}_j^{+,-,z}$, $\eta$ is the
classical drive amplitude $\eta=\sqrt{\frac{\kappa P_{in}}{\hbar \omega}}$, with $P_{in}$ being the input optical power, and $\hat{a}$ is the annihilation operator
of the cavity mode. The atom-cavity coupling rate is given by:
\begin{eqnarray}
  g_j(t)= g_0  \cos(kz_j-\delta_jt) e^{-r_j^2/w_0^2}\,,
\end{eqnarray}
where $k$ is the wave number of the cavity, $z_j$ and $r_j$ are the longitudinal and axial positions, $\delta_j=kv_j$ is the Doppler shift in terms of the atom velocity $v_j$, $w_0$ is the waist of the gaussian cavity mode, $g_0 = \wp/\hbar\sqrt{\hbar \omega_c/2\varepsilon_0 V_{\textrm{eff}}}$ is the vacuum Rabi frequency with $V_{\textrm{eff}}$ the effective mode volume of the cavity, $\wp$ is the dipole moment of the atomic transition, and $\varepsilon_0 $ is the vacuum permittivity

The incoherent evolution is described by the Liouvillian
$\hat{\mathcal{L}}\left[ \hat{\rho} \right]$,
\begin{eqnarray}
  \hat{\mathcal{L}}\left[ \hat{\rho} \right] &=&
  -\frac{\kappa}{2} \left\{ \hat{a}^{\dagger} \hat{a} \hat{\rho}
    + \hat{\rho}  \hat{a}^{\dagger} \hat{a}
    - 2 \hat{a} \hat{\rho} \hat{a}^{\dagger} \right\}
  \nonumber
  \\
  &&-\frac{\gamma}{2} \sum_{j=1}^N \left\{ \hat{\sigma}_{j}^{+}
    \hat{\sigma}_{j}^{-} \hat{\rho}
    + \hat{\rho} \hat{\sigma}_{j}^{+} \hat{\sigma}_{j}^{-}
    - 2  \hat{\sigma}_{j}^{-} \hat{\rho} \hat{\sigma}_{j}^{+}
  \right\}
  \nonumber
  \\
  &&+\frac{1}{2T_2} \sum_{j=1}^N \left\{ \  \hat{\sigma}_{j}^{z}
    \hat{\rho}  \hat{\sigma}_{j}^{z} -  \hat{\rho}
  \right\},
\end{eqnarray}
where $\kappa$ is the decay rate of the cavity, $\gamma$ is the
spontaneous emission rate for the atoms, and $1/(2T_2)$ is the
inhomogeneous dephasing.

We derive $c$-number Langevin equations corresponding to equation
(\ref{ME1}). Assuming that the classical drive $\eta$ is strong enough
a mean-field description provides an accurate representation. We
therefore define the mean values $\alpha = i \left< \hat{a} \right>,
\sigma^{-}_{j}= \left< \hat{\sigma}^{-}_{j} \right>, \sigma^{+}_{j}=
\left< \hat{\sigma}^{+}_{j} \right> , \sigma^{z}_{j}= \left<
  \hat{\sigma}^{z}_{j} \right>$ which evolve according
to
\begin{eqnarray}
  \dot{\alpha}&=&-\kappa \alpha + \eta +
  \sum_{j=1}^{N} g_j(t) \sigma^{-}_{j},
\label{a1}
\\
\dot{\sigma}^{-}_{j} &=&-\left( \frac{1}{T_2}
  + i \Delta \right) \sigma^{-}_{j} +  g_j(t) \alpha \sigma^{z}_{j},
\label{sm1}
\\
\dot{\sigma}^{z}_{j} &=& -\gamma \left( \sigma^{z}_{j}
  + 1 \right) - 2  g_j(t) \left( \alpha \sigma^{+}_{j}
  +  \alpha^{*} \sigma^{-}_{j} \right).
\label{sz1}
\end{eqnarray}

In the moving frame of reference of each atom, the cavity field
appears as a travelling wave, containing two frequencies shifted above
and below the cavity frequency by the Doppler shift. To solve this
problem that intrinsically contains a bi-chromatic drive we proceed in
two ways.

We first numerically integrate equations (\ref{a1}) - (\ref{sz1}),
approximating the sum in equation (\ref{a1}) as an integral, with the integrand weighted by the thermal velocity distribution. The integral is then partitioned into finite segments. The positions are chosen in an analogous manner, with the atoms being distributed in a gaussian distribution of  width $2w_0$ by the MOT. The velocity partition must be chosen with care, since the system exhibits Doppleron resonances. Specifically, at lower velocity, more resolution in the partition is required.

We adopt a floquet analysis \cite{PhysRevA.48.3092}, in which we define for each atom,
 \begin{eqnarray}
\sigma^{-}&=& \sum_l e^{i l \delta t} x_1^{(l)},
\nonumber
\\
\sigma^{+}&=& \sum_l e^{i l \delta t} x_2^{(l)},
\nonumber
\\
\sigma^{z}&=& \sum_l e^{i l \delta t} x_3^{(l)},
\nonumber
\end{eqnarray}
where $l=2p$ represents the contribution from Doppleron resonance of order $p$, and $p+1$ is the number of photons absorbed from one direction and $p$ photons emitted into the opposite direction as stated in the article. Upon substitution into equations (\ref{a1} -\ref{sz1}), equations for the amplitudes $x_1^{(l)}$,$x_2^{(l)}$, and $x_3^{(l)}$ are found:

\begin{eqnarray}
\dot{x}^{(l)}_{1} &=&-\left(i (\Delta+l \delta) + \frac{1}{T_2} \right) x^{(l)}_{1} +  \frac{\alpha}{2} \left( \beta x^{(l+1)}_3 + \beta^{*} x^{(l-1)}_3 \right),
\label{x1}
\nonumber
\\
\\
\dot{x}^{(l)}_{2} &=&\left(i (\Delta+l \delta) - \frac{1}{T_2} \right) x^{(l)}_{2} +  \frac{\alpha^{*}}{2} \left( \beta^{*} x^{(l+1)}_3 + \beta x^{(l-1)}_3 \right),
\label{x2}
\nonumber
\\
\\
\dot{x}^{(l)}_{3} &=& -\gamma \delta_{l,0} - \left( i l \delta + \gamma \right) x^{(l)}_{2}
\nonumber
\\
&&-\left(  \beta \alpha x^{(l+1)}_2 + \beta \alpha^{*} x^{(l+1)}_1 +  \beta^{*} \alpha x^{(l-1)}_2 + \beta^{*} \alpha^{*} x^{(l-1)}_1 \right),
\label{x3}
\nonumber
\\
\end{eqnarray}
where
\begin{equation}
\beta=g_0  e^{i k z} e^{-\frac{r^2}{w_0^2}}.
\end{equation}

For a given $\alpha$, Equations (\ref{x1} - \ref{x3}) define a linear system of equations, which are solved by truncating $l$ at some finite value, and inverting the system.

As a second method, we consider the steady state solution where equation (\ref{a1}) becomes,
\begin{equation}
0 = -\kappa \alpha+ \eta +\frac{g_0 N}{2} \int_{-\infty}^{\infty} d\delta P(\delta) \left( \beta x^{(-1)}_1 +\beta^{*} x^{(1)}_1  \right),
\label{anm}
\end{equation}
where the sum over atoms  has again been approximated as in integral, and $P(\delta)$ is the thermal velocity distribution.
The $\alpha$ that solves equation (\ref{anm}) is found by applying Newton's method for root finding. We see excellent agreement between the two methods.

\subsection*{Effect of Reduced Temperature}
\begin{figure}
\includegraphics[width=0.45\textwidth]{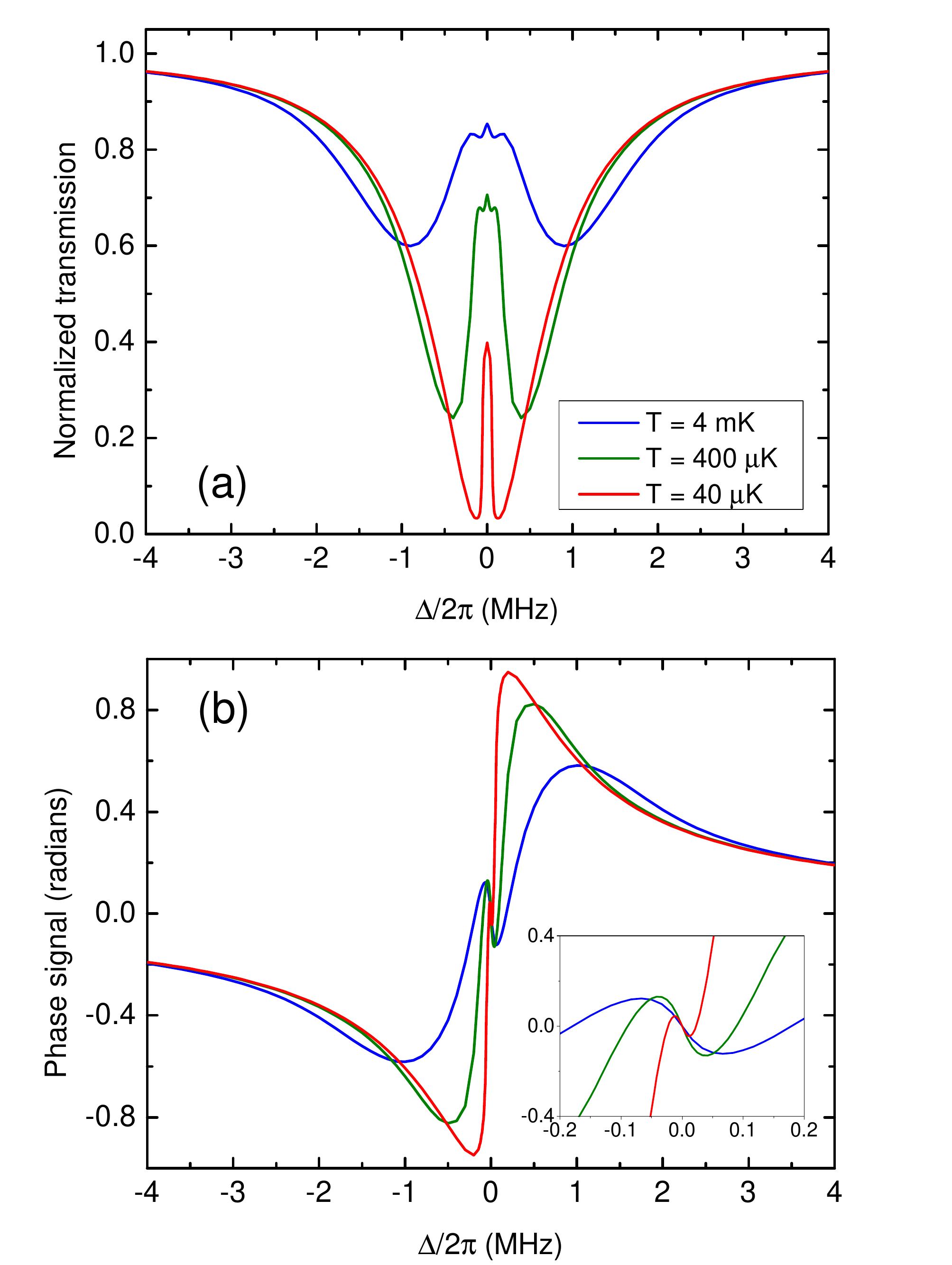}
\caption{\small (color online). Theoretical plots of the variation of the transmission (a) and the phase (b) with the temperature of the atoms. The input power and number of atoms are the same as in Fig. 2 in the article. The temperature $T=4$ mK corresponds to a typical experimental situation. The inset in (b) shows a zoom on the central part of the plot with the same units on the axes.
\label{fig:Temp_Dependence}}
\end{figure}

It is interesting to compare the situation observed in Fig. 2(b) and Fig. 2(c) in the article against lower temperature results. In Fig. \ref{fig:Temp_Dependence} we plot theoretical transmission and phase curves for $T=4$ mK,  $T = 400\; \mu$K and $T = 40\; \mu$K, where $T= 4$ mK corresponds to the typical experimental situation. The effect of going to lower temperature is evident on the transmission curves, where the Dopplerons become less and less dominant. The phase is less affected by Dopplerons, although the slope on resonance tends to increase slightly with decreasing temperature. The slope around zero-frequency determines the potential stability of this system in application as a frequency-lock. However, considering the additional experimental complexity of decreasing the temperature by orders of magnitude and the possible reduction of atom number in the process, the gain in slope is minimal.

\subsection*{Shot Noise Limited Linewidth}
Based on our experimental parameters we have estimated the shot noise limited locking performance of our setup. We proceed as in \cite{MikeMartin} to determine the shot noise limited linewidth with one difference: we do not assume the local oscillator power, \emph{i.e.} the sum of the power in each sideband $P_{\textrm{sideband}}$, is large compared to signal power $P_{\textrm{carrier}}$. $P_{\textrm{carrier}}=2 \kappa h \nu |\alpha|^2$ is the carrier power leaking out of the cavity.
Generally, when using the NICE-OHMS detection scheme, the local oscillator power may be comparable to the sideband power. Lifting the high local oscillator power assumption modifies the result by a factor of $\left(1+\frac{P_{\textrm{carrier}}}{2P_{\textrm{sideband}}} \right)$, resulting in
\begin{eqnarray}
\Delta \nu = \frac{\pi h \nu}{2 P_{\textrm{carrier}} \left(\frac{d \phi}{d\nu} \right)^2} \left(1+\frac{P_{\textrm{carrier}}}{2P_{\textrm{sideband}}} \right). \label{eq:linewidth}
\end{eqnarray}
Our calculation assumes unit quantum detector efficiency, and additional quantum noise due to fluctuations in intra-cavity photon number is expected to be unimportant (a $\sqrt{n}$ effect where $n$ is the number of photons in the cavity). The intra-cavity power and the phase slope $(\frac{d \phi}{d \nu})$ are calculated using the Floquet simulation, and found to be in excellent agreement with experimental values as shown in the article. For our present numbers, $\left(1+\frac{P_{\textrm{carrier}}}{2P_{\textrm{sideband}}} \right) = 30$, and we estimate a shot noise limited laser linewidth of $\Delta \nu = 1000$ mHz.

This linewidth estimate could realistically be improved by at least an order of 10-20 by improving the experimental parameters, thus reaching state-of-the-art performance.
First of all, optimizing the EOM modulation index to allow half of the optical power in the sidebands, such that $\left( 1+\frac{P_{\textrm{carrier}}}{2P_{\textrm{sideband}}} \right) = 2$, would immediately decrease the linewidth by a factor 15. Secondly, the phase slope scales directly with atom number, and an increase in the number of atoms by a factor of 3-5 is realistic (\emph{i.e.}, a decrease of the linewidth by a factor 9-25, see equation \eqref{eq:linewidth}). Possible atom numbers exceeding even $10^9$ have been reported in literature \cite{Stellmer} by operating the strontium oven at somewhat higher temperatures compared to our setup and employing more efficient Zeeman slower design in combination with strong transverse cooling section. Efficient transverse cooling alone may increase number of trapped atoms number by a factor of 4.
Additionally, increasing the finesse of the cavity a factor of 10, from the current $F = 85$ to $F = 850$, would reduce the laser linewidth by a factor of 100. However, to avoid undesirable power broadening effects the input power here must also be lowered. This makes the experimental realization of this method less attractive as it adds more technical challenges on the detection system. A realistic linewidth reduction due to an increase in the finesse we estimate as a conservative factor of 10.

\begin{figure}[h]
\includegraphics[width=0.45\textwidth]{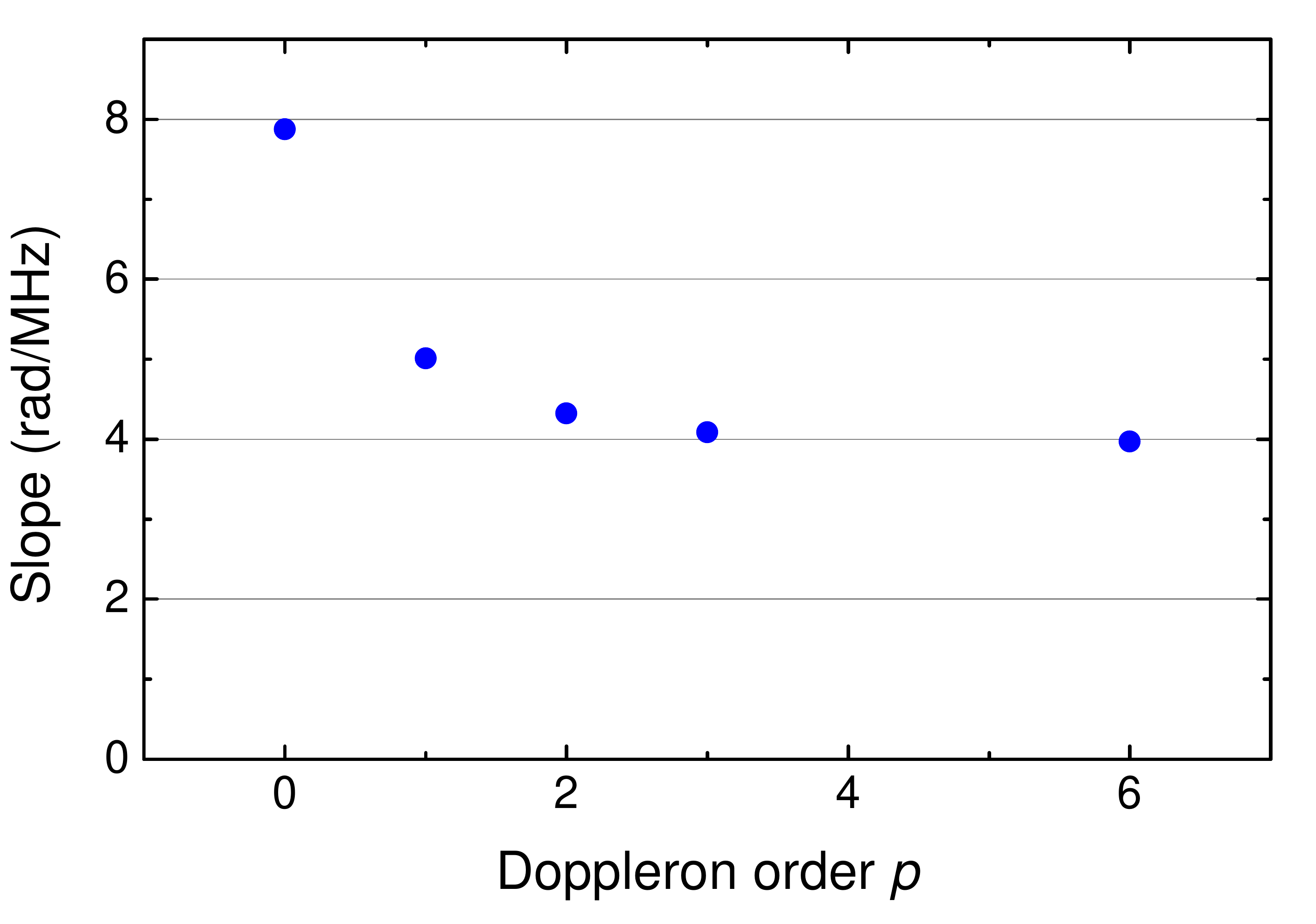}
\caption{\small (color online). Phase slope $\left(\frac{d \phi}{d \nu} \right)(p)$ on resonance (zero detuning) as a function of Doppleron order $p$. Here we sum all contributions to the slope with orders up to $p$. For $p$ = 0, only the zeroth order slope is included, while for $p$ = 1 both the 0 and the 1 order contribution to the slope are included and so forth.
\label{fig:SlopeDopplerons}}
\end{figure}

From the Floquet theory we may further investigate the role of the Doppleron resonances on the shot noise limited linewidth. This is done by decomposing the slope $\left(\frac{d \phi}{d \nu} \right)(p)$ according to the Doppleron order $p$. In the Floquet decomposition the order $\mathit{l}$ corresponds to the $p=\mathit{l}$/2 order Doppleron resonance \cite{PhysRevA.48.3092}, where $p=0,1,2\ldots$. To investigate the importance of Dopplerons, we compare the slope of the phase around resonance with increasing number of Doppleron orders included. To make the comparison, we use the parameters of Fig. 2 from the article, i.e., an input power of 975 nW, a total number of atoms in the MOT of $N= 4.4\cdot 10^8$ and temperature $T= 2.3$ mK. The slope for varying orders of Dopplerons is shown in Fig. \ref{fig:SlopeDopplerons}, where we sum all contributions to the slope $\left(\frac{d \phi}{d \nu} \right)(p)$ with orders up to $p$.

We find that the solution is sufficiently converged by third order, i.e., the sum of contributions from $p=0,1,2,3$. The difference between third order and sixth order solutions is less than a few percent. The slope change is found to be $\left(\frac{d\phi}{d\Delta} \right)_{p=0,1,2,3}/\left(\frac{d\phi}{d\Delta} \right)_{p=0} = 0.5$ at resonance for our conditions. As the shot noise limited linewidth scales as $\left(\frac{d \phi}{d \nu} \right)^{-2}$ the predicted linewidth calculated with Dopplerons will be larger than the linewidth calculated without Dopplerons by a factor of 4.

\end{document}